**Type of manuscript:** Original Contribution

**Manuscript title:** "Classification of spatiotemporal data identifies limitations of epidemic alert systems: Monitoring influenza-like illness in France"

**Running head:** "Spatial heterogeneity and epidemic alert systems"

**Author's names and affiliations:**

Pavel Polyakov[1], Cécile Souty[2], Pierre-Yves Böelle[2] and Romulus Breban[1]

[1]*Institut Pasteur, UEME, F-75015 Paris, France*

[2] *Sorbonne Université, INSERM, Institut Pierre Louis d'Epidémiologie et de Santé Publique, APHP, Hôpital Saint-Antoine, F-75012 Paris, France*

*F-75013 Paris, France*

**Corresponding author:**

Romulus Breban, PhD

phone: +33 (0)1.40.61.39.65

fax: +33 (0)1.45.68.88.76

email: romulus.breban@pasteur.fr

**ABSTRACT**

Surveillance data serving for epidemic alert systems are typically fully aggregated in space. However, epidemics may be spatially heterogeneous, undergoing distinct dynamics in distinct regions of the surveillance area. We unveil this in retrospective analyses by classifying incidence time series. We use Pearson correlation to quantify the similarity between local time series and then classify them using modularity maximization. The surveillance area is thus divided into regions with different incidence patterns. We analyzed 31 years of data on influenza-like-illness from the French system Sentinelles and found spatial heterogeneity in 19/31 influenza seasons. However, distinct epidemic regions could be identified only 4-5 weeks after the nationwide alert. The impact of spatial heterogeneity on influenza epidemiology was complex. First, when the nationwide alert was triggered, 32-41% of the administrative regions were experiencing an epidemic, while the others were not. Second, the nationwide alert was timely for the whole surveillance area, but, subsequently, regions experienced distinct epidemic dynamics. Third, the epidemic dynamics were homogeneous in space. Spatial heterogeneity analyses can provide the timing of the epidemic peak and finish, in various regions, to tailor disease monitoring and control.

Syndromic surveillance systems (1-6) are operational in many countries (7-18). Such systems routinely gather data on the number of clinical cases of infectious diseases over large surveillance areas. After curation and consolidation, data can be summarized as time series of disease incidence. Nationwide epidemic alert systems (19-23) are typically based on aggregating data over the whole surveillance area (i.e., highest level of spatial aggregation), to inform top-down strategies of public health. However, epidemics may be highly heterogeneous in space. Therefore, alerts from global systems may not concur with local epidemic dynamics; the national public health message may arrive either too late or too early for the local health authorities.

In this work, we develop new analyses of surveillance data to assess spatial heterogeneity of epidemics, using elements of metapopulation theory (24). In this context, the problem of disentangling the dynamics occurring in a large surveillance area is known as the *mega-patch problem* (25). Specifically, our problem is to classify large-scale, spatiotemporal surveillance data and figure out weakly interacting epidemics of the same infectious disease in subpopulations inhabiting different demographic areas. Hence, we aim to assess epidemic heterogeneity in the surveillance area and aggregate the data by subpopulation, to analyze the performance of global epidemic alert systems.

We apply our methods to influenza-like illness (ILI) data collected in the metropolitan France by the Sentinelles surveillance system (26, 27), a network of voluntary, unpaid, general practitioners who report weekly numbers of ILI diagnostics together with age, sex, vaccination status and clinical characteristics of patients. Influenza-like illness was defined as a sudden onset of fever over 39 degrees Celsius with myalgia and respiratory symptoms (cough, sore throat) (28).

Time series for the number of ILI cases per 100 000 individuals per week (i.e., ILI incidence), from 1984 to present day, are available at four different levels of spatial resolution: 96 departments (NUTS3), 22 administrative regions (NUTS2), 13 administrative regions (NUTS1) and the national level. The variance of incidence is estimated assuming that the number of reported cases obeys the Poisson statistics. Hence, 95% confidence intervals (CI) are calculated for all weekly incidence estimates, using a normal approximation (29). Furthermore, a weekly threshold for the national alert of influenza epidemic is estimated from the ILI incidence time series (18). The nationwide alert for influenza epidemic is triggered when ILI incidence exceeds this threshold for two consecutive weeks (18). Demographic data matching the time period of the Sentinelles data are available from the Institut National de la Statistique et des Etudes Economiques (30).

**METHODS**

We assume that the surveillance area is divided into *N* smaller units for simultaneous data collection, called *catchment areas*. In each catchment area, data is collected independently and may be further used to reconstruct a time series of incidence, to be assigned to the catchment area.

Epidemics occur in spatially heterogeneous demographic environments. Hence, the epidemic dynamics at the national scale may originate from a sum of several, say $C$, nearly independent epidemics, each of them established into a distinct community, included in the surveillance area. A community epidemic may be localized in a region spanning several catchment areas, called *epidemic region*. See Fig. 1, where catchment areas $i$ and $j$ are included in epidemic region $a$, while catchment area $k$ is included in epidemic region $b$. Studying interdependence between incidence time series, constructed for each catchment area, may suggest how to group/classify these space units and reconstruct the epidemic regions (Fig. 1).

Over a short time interval, the total number of clinical cases reported in a catchment area, comprised into an epidemic region, has two contributions. It contains (1) a fraction of the infectious disease cases reported in the epidemic region and (2) other cases, which may not belong to the epidemic, and represent a time-independent background, specific to the catchment area. Hence, the incidence of catchment area $i$, $I_i$, may be written as a fraction $p_i$ of the infectious disease incidence in epidemic region $a$, $J_a$, supplemented by a local background incidence $q_i$; i.e., $I_i = p_i J_a + q_i$. Thus, time series from two catchment areas included in the same epidemic region fit well one versus the other through a linear model, while time series from two catchment areas, not included in the same epidemic region, do not. The purpose of our methodology is to estimate the number of epidemic regions and identify the catchment areas included in each epidemic region (Fig. 1).

The analyses are organized according to the following steps:

(i) *Assessment of time-series interdependence.* We consider each pair of catchment areas and fit their corresponding time series, one versus the other, using a linear model. In case of missing data, we select only the data at common time points. The adjusted square coefficient of determination, $R^2$, used as goodness of fit, characterizes the correlation strength between each pair of time series. $R^2$ is also proportional to the square coefficient of Pearson correlation and is thus symmetric in the two time series, defining a reflexive relationship of similarity. Hence, the $R^2$ values from all pairs of time series can be organized as a symmetric matrix.

(ii) *Determination of epidemic regions.* We consider the matrix of $R^2$ values as an adjacency matrix, defining an all-to-all undirected network with weighted edges, having all catchment areas as nodes. We classify the time series corresponding to the catchment areas by determining the community structure of this all-to-all network using modularity maximization (31-34). Thus, each community contains nodes (catchment areas) whose time series correlate well to one another and forms an *epidemic region*. In contrast, the correlation between time series of catchment areas included in different communities (epidemic regions) may be significantly weaker. Modularity maximization is employed here as an algorithm for unsupervised classification or hierarchical clustering (34-36), using a measure of similarity (i.e., $R^2$). Many other classification schemes are available. However, modularity maximization is simple to implement numerically and its methodological shortcomings are well documented (37).

We tested two numerical algorithms (31, 33) to maximize modularity and obtained similar results for our ILI data. The output of a single run of modularity maximization is a partition of the nodes into a number of disjoint communities, determined by the algorithm. In fact, each node is assigned a numerical label, from 1 to $C$, for the community where it belongs. The output is organized as a vector of community labels, with length equal to the number of nodes. This data structure, fundamental for our analyses, is called *modularity vector* (31, 33).

(iii) *Impact of data uncertainty.*

Epidemic regions result from classifying a relatively small numbers of nodes (catchment areas) whose features are defined by noisy data. Bootstrap analyses can establish the robustness of the node classification with respect to noise. Here, we organized according to the following three steps.

(iiia) *Surrogate datasets.*

We built surrogates for each data point using a normal distribution with mean at the incidence estimate and variance tuned by the corresponding confidence interval (i.e., parametric bootstrap). We thus generated 10 000 surrogate datasets, which we analyzed according to steps (i) and (ii) described above.

(iiib) *Bootstrap analyses.*

The labeling of the communities serves no particular purpose beyond single run analysis. This leads to difficulties in the case where many modularity vectors are to be processed as an ensemble. Even in the case where nearly similar datasets are processed, all yielding exactly the same partition, the community labeling may differ, leading to different modularity vectors. However, in this case, we can align each modularity vector to the first, by searching among all permutations of symbols from 1 to $C$ for a relabeling of the communities that makes each modularity vector coincide with the first.

The same principle works for the case where datasets are not too similar (still, noise in the data should be sufficiently small), so the resulting modularity vectors correspond to different, but similar partitions. In general, analysis of bootstrap data yields an ensemble of modularity vectors where the number of communities, $C$, ranges from 1 to some maximum value, $C_{max}$. For each value of $C$, we chose a reference vector, which we align as follows. Given two modularity vectors with $C_1$ and, respectively $C_2$ communities ($C_1 \leq C_2$), we search among all permutations of symbols from 1 to $C_1$ for a relabeling that maximizes the number of identical symbols between the two vectors. The remaining (i.e., non-reference) modularity vectors may be then quickly aligned to the reference vector with the corresponding value of $C$, using a similar procedure. Alignment of modularity vectors ensures the same community labeling for the analyses of all bootstrap data.

(iiic) *Summary statistics for the ensemble of aligned modularity vectors*

The ensemble of aligned modularity vectors is further processed to reveal distinctions in the node classification based on details attributable to noise in the data. In particular, we compute $f_a^i$, the fraction of times catchment area $i$ appeared in community $a$. To evaluate the global impact that uncertainty has on community structure, we calculate the following bootstrap score

$$B = -\sum_{a=1}^{C_{max}} \sum_{i=1}^{N} \frac{f_a^i \log(f_a^i)}{N \log(N)}.$$

If all values of $f_a^i$ are either 0 or 1, then data uncertainty has no impact on community decomposition and $B=0$; the higher the value of $B$, the deeper the impact of data uncertainty on community decomposition; see section 1 in the supporting document (SD) for further discussion.

(iv) *Data aggregation for each epidemic region*. Incidence time series belonging to community $a$ may be aggregated using the catchment area population multiplied by $f_a^i$, as weight. This yields a time series of expected incidence and corresponding CI for each community. We do not aggregate a time series for community $a$ if $\sum_{i=1}^{N} f_a^i < 2$; that is, the weights for region $a$ do not amount to represent at least two catchment areas.

(v) *Validation*. We reject the analyses if the resulting bootstrap score $B$ is above a pre-established threshold. Furthermore, we use Hellinger distance (38) to validate statistically the distinction between two aggregated time series. Namely, for each moment of time, we calculate the Hellinger distance between the corresponding data, assuming that incidence is normally distributed. Then, summing the pairwise Hellinger distances over the duration of the time series, we obtain a Hellinger score, denoted by $H$. The distinction between two epidemic regions is rejected if the Hellinger score between the corresponding aggregated time series is less than a certain threshold. In this case, one may choose to merge the two epidemic regions, as the distinction revealed by the algorithm may be too small to have operational value.

The interdependence between aggregated, community time series may be further assessed using Pearson correlation; see step (i). Community decompositions with high bootstrap score $B$, where many values of $f_a^i$ differ significantly from 0 or 1, yield high $R^2$ values between aggregated time series, because the time series for catchment area $i$ may contribute significantly to several aggregated community time series. This is where our analyses do poorly. In contrast, for community decompositions with low bootstrap score $B$, we may expect that $R^2$ between aggregated time series is small, indicating weak interdependence and a meaningful disaggregation of the national epidemic curve. We may also conclude that our analyses are successful if the bootstrap score $B$ is low and, on average, only one community is found. In this case, we say that the entire surveillance area was subject to an epidemic homogeneous in space.

In the case of automated surveillance, where the lengths of the time series increase steadily with time, the above analyses may be repeated for each additional time point, to constantly monitor $B$, $H$ and $R^2$.

**RESULTS**

Each seasonal epidemic may have unique dynamic and spatial pattern. Hence, we performed our analyses for yearly datasets. Furthermore, we split yearly data into two: a 26-week period (the *influenza season*, mid-October to mid-April) when influenza epidemics can occur in France, and its yearly complement.

Analyses were carried out independently. During an influenza season, the weekly number of ILI cases per 100 000 individuals typically peaks at high values (i.e., 350-2000), while off-season, it passes through much lower values (i.e., <100). Our analyses showed that findings are robust regarding the definition of the influenza season, provided that the time interval defined by the start week (SD, Table S1) and the 26-week duration includes the high incidence values.

We contrasted analyses of data collected during the influenza season with analyses of data collected off-season (SD, Fig. S1). The adjusted $R^2$ values were relatively high for the data collected during the influenza season, motivating the search for epidemic regions (SD, Fig. S2). In contrast, $R^2$ values were consistently low for the data collected off-season, indicating that ILI incidence time series are poorly correlated. Thus, the search for epidemic regions was not motivated, in this case.

We performed analyses at 3 levels of spatial resolution (SD, Fig. S3). First we considered the department level (NUTS3; 96 departments). The typical French department had ~500 000 population and 3-5 practitioners participating to the Sentinelles surveillance system. Under the circumstances, fluctuations in reported ILI cases proved important for time series dynamics and trends were not readily apparent. The resulting $R^2$ values were relatively low and, searching for community structure at this level, unjustified (SD, Fig. S3). Analyses at the level of the largest regions (NUTS1; 13 regions) were successful. However, the spatial resolution was rather coarse. A better approach was the middle ground of spatial resolution, provided by the 22 administrative regions (NUTS2).

Bootstrap analyses unveiled the role of data uncertainty for our results (SD, Table S1). Based on retrospective analyses, we divided the 31 influenza seasons into four groups: seasons where (1) $B>0.1$ and data uncertainty had a significant impact on ascertaining spatial homogeneity/heterogeneity of epidemics (5/31), (2) $H<10$ and thus, aggregated time series were very similar, considering data uncertainty (5/31), (3) the level of spatial heterogeneity may be subject to further discussion (11/31), and (4) spatial heterogeneity was particularly strong, playing a clear role for surveillance (10/31). Group (4) is described in Table 1.

Retrospective analyses revealed three scenarios how epidemic heterogeneity played a role relative to the current protocols for epidemic alert and epidemic threshold in France. Accordingly, we present detailed results about six illustrative influenza seasons. Figure 2 shows results for the 2015-2016 and 1985-1986 influenza seasons, where the surveillance area of Sentinelles was divided into two epidemic regions of comparable size. We colored a catchment area depending on the fraction of times it occurred in the first epidemic region and the fraction of times it occurred in the second epidemic region, in bootstrap analyses of 10 000 datasets. A catchment area was colored blue or red if it occurred always in the first or second epidemic region; i.e., the fractions were, respectively, either (1.0, 0.0) or (0.0, 1.0). Otherwise, we used intermediate colors, linearly interpolated in the RGB palette. However, in the case of low $B$, all map colors are nearly blue or nearly red.

The two epidemic regions that we found have distinct influenza dynamics; see bottom panels in Fig. 2. The epidemic in one region (blue) started early, while the epidemic in the other region (red) started 3-4 weeks later. The national epidemic alert (vertical line) was triggered by the early epidemic. Hence, for 3-4 weeks, one region was under alert, yet not experiencing an epidemic, because of the epidemic in the other region. In these cases, simply mapping the difference between the local and national-level ILI incidence at the time of the national epidemic alert, suggested clear patterns of spatial heterogeneity (SD, Fig. S4).

Figure 3 shows results for the 1997-1998 and 1991-1992 influenza seasons. Again, in each case, the surveillance area was divided into two epidemic regions of comparable size. The epidemics arrived nearly at the same time; the delay was only 1 week, which is the time resolution of Sentinelles. However, the subsequent epidemic dynamics were quite different in the two regions that might have benefited from tailored surveillance. For example, in the 1997-1998 influenza season, ILI incidence peaked 5 weeks apart in the two epidemic regions. This was not the case for the 1991-1992 influenza season. However, during the 1991-1992 influenza season, ILI incidence in one epidemic region (blue) reached below the epidemic threshold 5 weeks before the ILI incidence in the other epidemic region (red).

Finally, we report on the 2014-2015 and 2012-2013 influenza seasons (Table 1 and SD, Table S1), where the whole surveillance area appeared as a single epidemic region in 97% and 85% of the bootstrap analyses, respectively. These are remarkable cases of epidemic homogeneity in space. Consequently, the national alerts were timely for the entire surveillance area.

To assess their performance for real-time applications, we repeated our analyses with varying amount of data. Namely, we considered datasets from the beginning of the influenza season up to a certain week after the epidemic alert and investigated how $B$ and $R^2$ change with the amount of data. Figure 4 shows results for the 2015-2016 and 1997-1998 influenza seasons, where the most likely divide was into two epidemic regions. The amount of data gathered from the beginning of the influenza season up to the time of the national alert, and immediately after, was insufficient for clear results. $B$ values were high, indicating that data uncertainty had a strong impact on our results. Furthermore, aggregated time series were highly correlated, resembling the national epidemic curve. However, data gathered 4-5 weeks after the nationwide alert was already sufficient to ascertain spatial heterogeneity of the epidemic with $B<0.1$ and $R^2<0.6$. Analyses of datasets for the following weeks provided confirmation, with a slightly increasing trend for $R^2$ as the epidemic went extinct.

It is important to note that, even if detection of epidemic regions may occur later than the nationwide alert, it can still offer critical information on the qualitative and quantitative dynamics of regional epidemics. For example, in the case of the 2015-2016 influenza season (Fig. 4), robust detection of epidemic regions is obtained in week 5 after the global epidemic alert. Inspection of Fig. 2 reveals that this corresponds to week 10 of the influenza season. At that time, the epidemic was past its peak in the blue epidemic region, but not in the red region.

Incidentally, the epidemic was also not past its peak at the national scale. Similar results hold for the 1997-1998 influenza season. Hence, our methods can successfully deliver the spatial structure of influenza epidemics, in time for improving monitoring and control during the influenza season.

**DISCUSSION**

Timeliness is a key performance measure of public health surveillance systems (39). However, timeliness can depend on the scale at which information is aggregated to inform public health practitioners. For example, in France, epidemic alerts for influenza are given once for the whole surveillance area, based on data aggregated at the national level. There are obvious statistical advantages in aggregating time series at the national level: noise is much reduced and incidence trends are more prominent as they cross epidemic thresholds. However, in case of spatial heterogeneity, this approach masks distinct epidemic dynamics taking place in the surveillance area, and undermines the usefulness of the surveillance system.

The case of influenza illustrates this problem perfectly. The burden of seasonal influenza epidemics is large in western countries (40). Once the epidemic alert is triggered, national media campaigns on prevention are launched in the press, TV and radio, with messages on hygiene and vaccination. Timing can be critical, since vaccine-induced protection becomes effective about 2 weeks after vaccination (41). However, more concerning is the impact of influenza epidemics on the routine functioning of hospital wards (42). In France, influenza epidemic alerts put hospitals under stress, as emergency protocols may be activated (e.g., postponing non-urgent interventions, conscripting staff, freeing hospital beds). Untimely decisions in these situations may lead to inefficiency and incumbent costs to society.

It is therefore important to identify situations where a public health message, designed for national broadcast, can be later supplemented with customized information for the local practitioners. Mapping incidence data already gives important clues about epidemic heterogeneity (43). Furthermore, gravity/radiation models (44-48) could be employed for extensive spatiotemporal analyses, including data fitting. However, these models make strong assumptions on how two catchment areas must interact; i.e., gravity models assume that interaction declines with (square) distance, while radiation models assume that individuals spread like radiation fluxes. With the goal of classifying spatiotemporal data, we made minimal assumptions about the geographical structure, using concepts of metapopulation theory to discover the spatial structure of epidemics. Still, the epidemic regions that we found cluster, to a large extent, neighboring catchment areas.

Regions of distinct epidemic dynamics may occur for a variety of reasons. Factors known to influence the transmission of influenza are: susceptibility profile of the population, circulating strains, vaccine parameters and vaccination patterns, travel and daily commuting, school holidays, and the weather. In addition, the spatiotemporal coordinates of the individuals who happen to initiate the influenza epidemics (so-called *patients zero*) may also be very important. Therefore, an account for the spatial heterogeneity of influenza, starting from first principles, may be particularly challenging. Our strategy was to determine influenza epidemic regions directly, using surveillance data, by analyzing local epidemic dynamics provided by incidence time series.

Acknowledging spatial heterogeneity in a surveillance area and identifying epidemic regions may have important consequences for improving influenza monitoring and control. The operational value of epidemic regions depends on several key items. First, the spatial resolution must be appropriate (49). Here, the signal to noise ratio was too small for a meaningful analysis at the NUTS3 level, but sufficient for defining epidemic regions at NUTS2 and NUTS1 level. This is relevant for the structure of public health administration in France. Second, the time to identification of distinct dynamics must be compatible with staggered delivery of public health messages. Here, 4-5 weeks of data past the time of the epidemic alert were necessary to identify epidemic regions. Even with this delay, valuable regional updates can be passed to the local health authorities during the epidemic, to maintain vigilance in the affected territories.

The distinction of epidemic regions may be also relevant for collecting pathogen samples. Every influenza season, clinical samples from ILI cases are collected and analyzed, to establish the spectrum of circulating virus strains. In turn, this determines the composition of future influenza vaccines. Clinical samples should be collected from each epidemic region for a better characterization of the circulating influenza strains.

In conclusion, we proposed new methodology to detect spatial heterogeneity in disease surveillance data and discussed monitoring ILI in France. However, our methods are not tailored to influenza epidemiology and may be used for other case diseases.

**Conflict of interest:** There are no conflicts of interest.

**Sources of funding:**

This work was supported by Agence Nationale de la Recherche (Labex Integrative Biology of Emerging Infectious Diseases) in the form of a postdoctoral scholarship for PP.

**Data and computer code:** Data on the incidence of influenza-like illness in France is freely available from the Sentinelles website (**https://www.sentiweb.fr**). Demographic data for France is freely available from the INSEE website (**http://www.insee.fr**). The code that we developed for data analyses is freely available upon request.

**TABLE LEGENDS**

**Table 1**. Summary of results for influenza seasons where the bootstrap score $B$ was lower than 0.1 and the adjusted $R^2$ between aggregated time series was lower than 0.6 or one epidemic region was most likely to occur. The most likely divide was into two epidemic regions, with the exception of the 2014-2015 and 2012-2013 influenza seasons for which only one region was most likely to occur. The probability for the most likely number of regions to occur was larger than 0.85, in each case.

**FIGURE LEGENDS**

**Figure 1.** Conceptual representation of a surveillance system observing a spatially heterogeneous epidemic. The surveillance area (thick contour) is divided into *N* catchment areas represented as squares (e.g., *i*, *j*, *k*). Several (say *C*) nearly independent community epidemics may develop simultaneously in various regions of the surveillance area; hence we may distinguish several epidemic regions (e.g., *a*, *b*). An epidemic region may include several catchment areas; e.g., catchment areas *i* and *j* are included in epidemic region *a*, while catchment area *k* is included in epidemic region *b*. The incidence time series of area *i* fits well the incidence time series of area *j* through a linear model, as catchment areas *i* and *j* belong to the same epidemic region and sample the same epidemic dynamics. In contrast, the incidence time series of area *i* fits badly the incidence time series of area *k*, as catchment areas *i* and *k* belong to different epidemic regions. The purpose of our methodology is to sort out the incidence time series in all catchment areas and attribute each area to an epidemic region.

**Figure 2.** Analyses for the 2015-2016 and 1985-1986 influenza seasons. The left panels refer to 2015-2016, while the right panels refer to 1985-1986 (Corsica was excluded due to missing data). The upper panels represent the epidemic

regions that we found for France. A NUTS2 region (i.e., catchment area) was colored depending on the fraction of times it occurred in the first epidemic region and the fraction of times it occurred in the second epidemic region, in the analyses of 10 000 bootstrap datasets. Blue corresponds to fractions (1.0, 0,0) to occur in the first and second epidemic regions, respectively, while red corresponds to the fractions (0.0, 1.0). For the other cases, we used intermediate colors. The lower panels represent the ILI incidence time series aggregated over the epidemic regions (blue and red, respectively), the national ILI incidence time series (black) and the epidemic threshold (dotted line). Note that the nationwide alert (vertical line) was not timely for both epidemic regions.

**Figure 3.** Analyses for the 1997-1998 and 1991-1992 influenza seasons. The panel arrangement and color code are similar to that in Fig. 2. The left panels refer to 1997-1998 (Corsica was excluded due to missing data), while the right panels refer to 1991-1992. The upper panels represent the epidemic regions that we found for France. A NUTS2 region (i.e., catchment area) was colored blue (red) if it occurred all the time in the first (second) epidemic region. Intermediate colors represent NUTS2 regions that occurred in both epidemic regions. The lower panels represent the ILI incidence time series aggregated over the epidemic regions (blue and red, respectively), the national ILI incidence time series (black) and the epidemic threshold (dotted line). While the nationwide alert (vertical line) was appropriate, the epidemic regions experienced distinct dynamics and could have benefited from customized monitoring.

**Figure 4.** The bootstrap parameter $B$ and the adjusted $R^2$ between the aggregated time series of the first and second epidemic regions as a function of the amount of data: results for the 2015-2016 and 1997-1998 influenza seasons. Data to estimate $B$ and $R^2$ was considered from the beginning of the influenza season up to a certain week after the epidemic alert, indicated on the horizontal axis. Threshold values for $B$ and $R^2$ at 0.1 and 0.6, respectively, are shown as horizontal dashed lines. We note a marked drop in $B$ and $R^2$ **4-5** weeks after the epidemic alert.

**TABLES**

| Season | Start week | Bootstrap score, $B$ | Hellinger score $H$ between aggregated time series | Adjusted $R^2$ between aggregated time series | Expected fraction of catchment areas in each epidemic region[c] |
|---|---|---|---|---|---|
| 2015-2016 | 48 | 0.006 | 13.3 | 0.49 | 0.585; 0.414 |
| 2014-2015[b] | 46 | 0.035 | - | - | 0.976; 0.024 |
| 2012-2013[b] | 46 | 0.074 | - | - | 0.934; 0.066 |
| 2011-2012[a] | 48 | 0.100 | 10.2 | 0.60 | 0.492; 0.508 |
| 2010-2011 | 44 | 0.028 | 13.4 | 0.46 | 0.572; 0.428 |
| 1997-1998[a] | 50 | 0.014 | 12.9 | 0.34 | 0.492; 0.508 |
| 1996-1997[a] | 42 | 0.014 | 14.4 | 0.56 | 0.436; 0.564 |
| 1991-1992 | 42 | 0.062 | 12.3 | 0.53 | 0.618; 0.356; 0.026 |
| 1989-1990 | 41 | 0.000 | 14.8 | 0.43 | 0.500; 0.500 |
| 1985-1986[a] | 44 | 0.026 | 13.4 | 0.41 | 0.341; 0.649; 0.010 |

[a]Corsica was excluded from the analyses due to missing data.
[b]The most likely number of epidemic regions was 1.
[c]We exclude values lower than 0.001. One administrative region represents $1/22 \approx 0.045$ ($1/21 \approx 0.048$ when Corsica is excluded).

**Table**

**FIGURES**

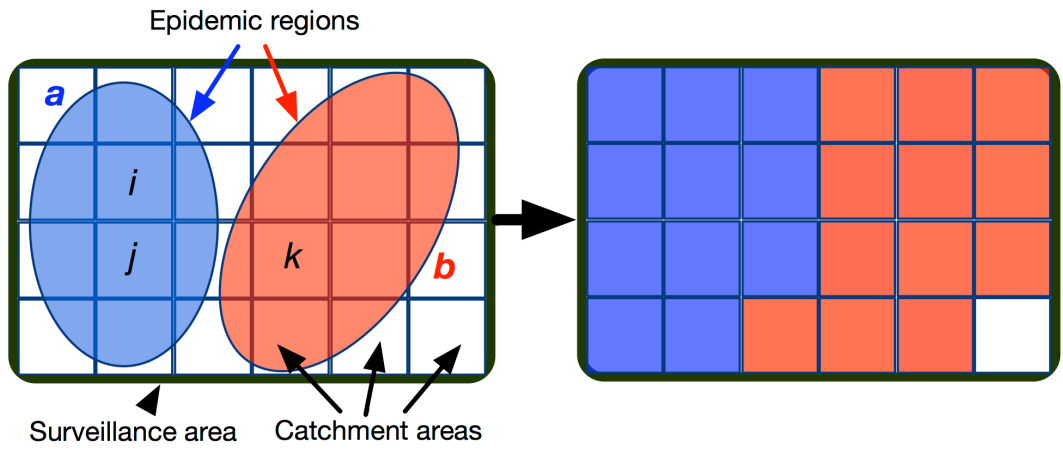

**Figure 1**

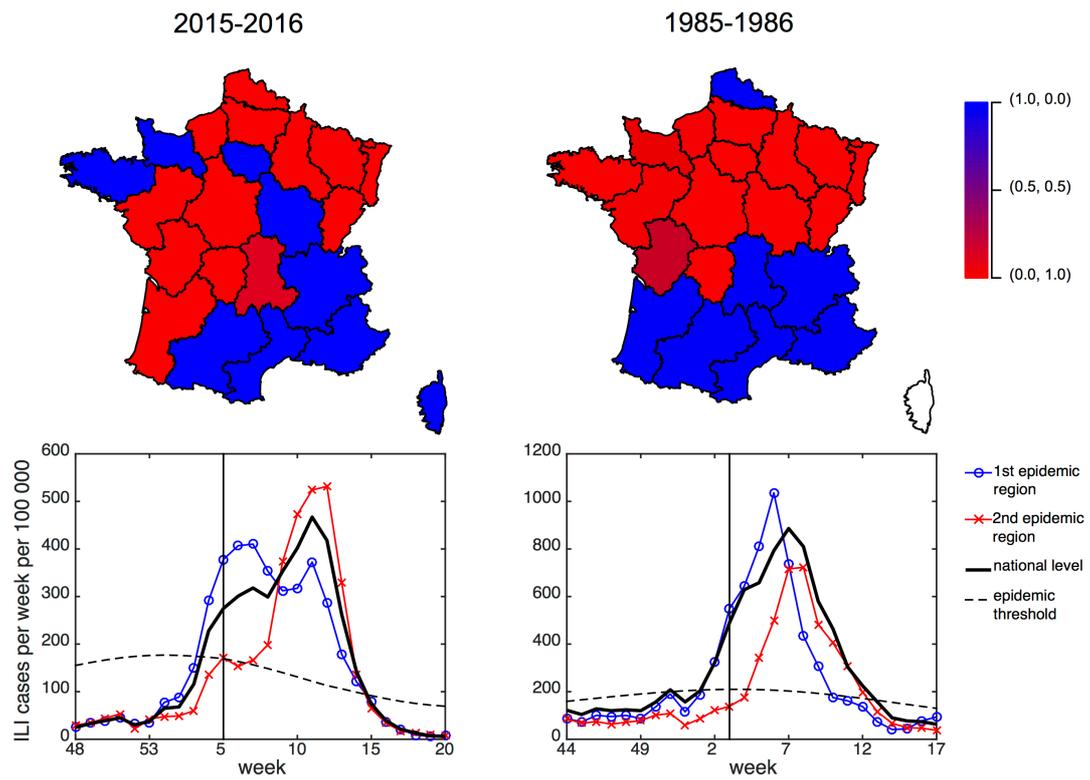

**Figure 2**

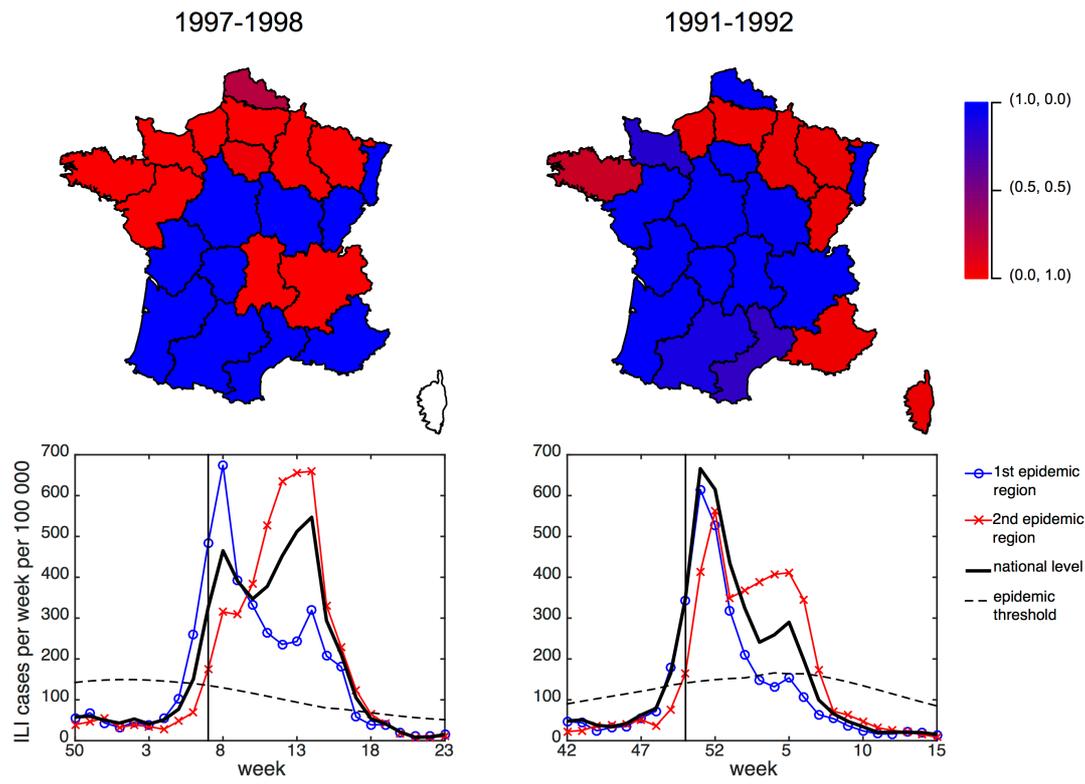

**Figure 3**

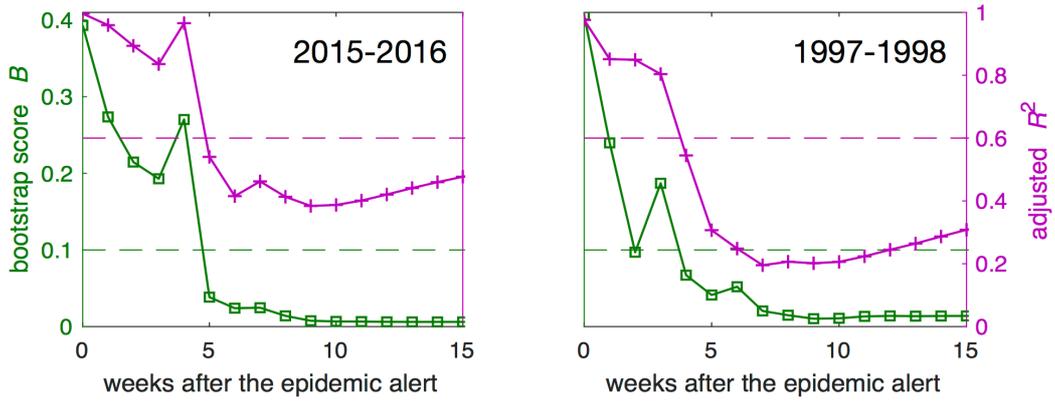

**Figure 4**